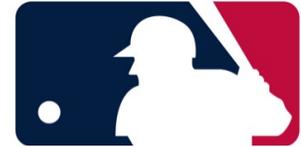

# You Cannot Do That Ben Stokes: Dynamically Predicting Shot Type in Cricket Using a Personalized Deep Neural Network

Will Gürpınar-Morgan, Daniel Dinsdale, Joe Gallagher,
Aditya Cherukumudi & Patrick Lucey
Track: Other Sports
Paper ID: 1548748

## 1. Introduction

The ability to predict what shot a batsman will attempt given the type of ball and match situation is both one of the most challenging and strategically important tasks in cricket.

The goal of each batsman is to score as many runs as possible without being dismissed. Batsmen can be dismissed in several ways, including being caught by fielders or having their wickets knocked over. While simple in principle, the type of shots and style of a batsman is greatly influenced by the format of the game. In short forms of the game such as T20 and One Day Internationals (the focus of this paper), batsmen are typically more aggressive since their team have a limited number of balls from which to score their runs (120 and 300 balls respectively).

Getting the right batsman vs bowler match-up is of paramount importance.  For example, for the fielding team, the choice of bowler against the opposition star batsman could be the key difference between winning or losing. Therefore, the ability to have a predefined playbook (as in the NFL) which would allow a team to predict how best to set their fielders given the context of the game, the batsman they are bowling to and bowlers at their disposal would give them a significant strategic advantage.

***In this paper, we present a personalized deep neural network approach which can predict the probabilities of where a specific batsman will hit a specific bowler and bowl type, in a specific game-scenario.***

As a motivating example let us consider the 2019 Cricket World Cup Final between England and New Zealand, with England needing 9 runs from 3 balls to win. The ball was an attempted "yorker" length delivery, affectionately known as a "toe cruncher" and renowned for being harder to hit long distances. However, the ball missed its mark and failed to bounce - a "full toss" length in cricketing terms - making it much easier to hit (see Figure 10-B for reference to the length nomenclature).  In Figure 1 we visualize how the predicted zone likelihood of Ben Stokes' shot type for this delivery varies using our deep learning model, where we vary the bowling length while holding all other aspects of its original trajectory fixed. The left plot shows the predicted shot location of the actual delivery; we then gradually decrease the length of the ball to the attempted yorker (center) and finally a short-pitched delivery (right). We can clearly see that the model predicts the outfield zone in the mid wicket region to be Stokes' preference for this line of delivery but the absolute magnitude changes by almost 10% and 20% when comparing the yorker with the full toss and short lengths respectively.

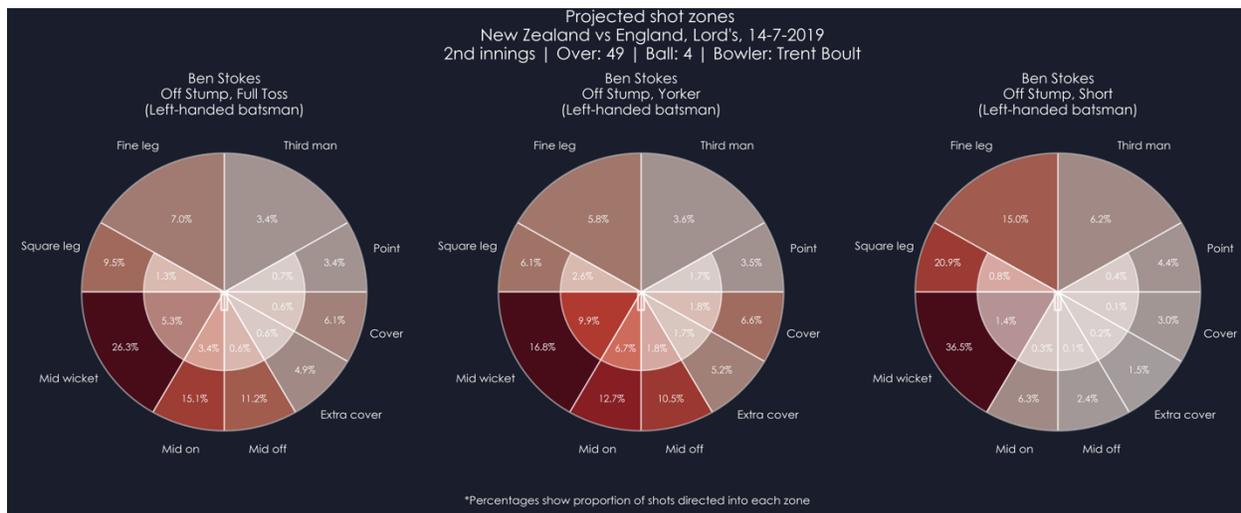

*Figure 1: Using our personalized deep neural network model we can predict various shot types based on the specific bowler, game-state and ball trajectory. Above shows the predicted shot charts for Ben Stokes with 3 balls to go against Trent Boult in the 2019 World Cup final for different length deliveries.*

The importance of this work is that it can be used across many different elements of the sport. First, for team performance applications, teams can create both pre-game strategies and in-game tactics that evolve throughout an innings as the match context changes. The wealth of information in our model will help teams plan bowling tactics and corresponding fielding locations. Secondly, for media where data drives many storylines for broadcasters, having an estimate of where players are most likely to hit shots in a given situation would enable deeper and more powerful storytelling, which would go beyond current score and win predictors which are currently utilized.

## 2. Background on Cricket

Although often viewed as a niche and somewhat impenetrable sport, cricket is more accessible and global than some realize; in fact, the first international cricket match did not involve teams which we typically associate with the sport and was between the USA and Canada on 24th September 1884 in New York. Since then, the sport has grown into one of the most popular and lucrative in the world, with over 100 member nations and huge TV audiences, such as India vs Pakistan in the 2019 World Cup which saw in excess of 250 million unique viewers [1]. A key indicator of its extensive commercial clout is reflected in the enormous rights associated with the Indian Premier League in 2019 at US$6.8 billion – despite its recency, this competition is already approaching the financial pull of the English Premier League in soccer [2].

The aim of cricket is simple: score more runs than the other team. Scoring runs is conducted in a similar manner to baseball, with one player bowling (pitching) the ball to a batsman who defends 3 wooden stumps (called wickets) and attempts to hit the ball in order to try and accumulate runs. However, unlike baseball, the legal hitting area in cricket is 360 degrees from the location of the batsman, who plays on a rectangular pitch in the centre of the playing field as demonstrated in Figure 10-A in the appendix. This large hitting area must be covered by 10 fielders, one of which is the wicketkeeper (equivalent to catcher in baseball) who typically stands directly behind the batsman.

Scoring runs can done in one of two ways. Firstly, by hitting boundaries, 4 runs if the ball clears the playing area along the ground and 6 runs, if it's hit over the playing area. Secondly - and most frequently - a batsman can score runs (1, 2 or 3) by swapping with their partner who stands at the opposite end of the pitch before the ball is returned to the stumps, similar to players running from one base to another in baseball.

Aggression brings risk however, as bigger shots carry a higher likelihood of dismissals such as being caught by fielders. Once 10 of a team's 11 batsmen have been dismissed their innings is complete, even if there are balls left to bowl. Therefore, a careful balance between aggression and caution is required to maximize the overall team score.

The direction and aggression with which a batter attempts to hit a delivery depends on several contextual factors that are a mixture of premeditated decision-making and split-second reactions once the ball is delivered: where the ball bounces, the speed and spin of the ball being delivered, and the current match state.

The goal of the fielding team is to dismiss the batsman and/or limit their run-scoring. The three main factors that the fielding team can control to increase the probability of a favorable outcome are:
   i.   The placement of fielders (subject to restrictions on the number of players patrolling specific regions)
   ii.  The choice of bowler at a given stage of a match (subject to a maximum number of balls per bowler)
   iii. The speed and trajectory of the ball (subject to the skill and consistency of the bowler)

What the fielding team cannot control at the moment of the delivery is the match context, atmospheric/weather conditions and factors intrinsic to the batsman, such as their level of skill or decision-making. However, the fielding team can leverage their understanding of the ability and tendencies of a batsman to restrict their impact.

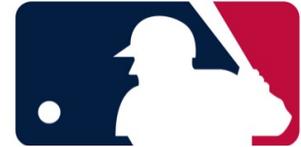

Bowling teams must attempt to process all of this information on the field in real-time to plan their strategies; fundamentally they are trying to predict what type of shot a batsman will play and where they will direct it. As a prediction problem, this is incredibly challenging due to the breadth and depth of cricket – with its 360-degree nature and wide array of shot types that can range from the brute strength required to launch the ball high into the stands, through to deft touches that barely alter the trajectory of the ball as it rolls gently across the grass. All batsmen have their particular strengths, weakness and preferences for shot type, which can drastically alter the expected shot direction for any given delivery. Contrast this with baseball, where the hitting arc is 90 degrees and handedness of the batter hugely narrows the optimum hitting angle further.

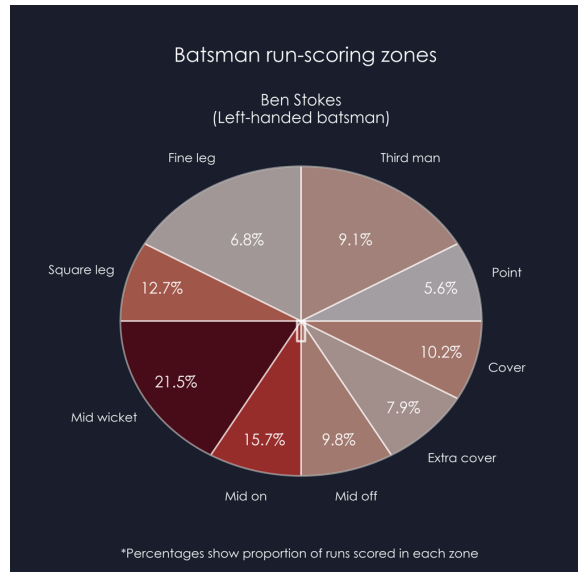

*Figure 2: Proportion of runs scored in various zones of the field by Ben Stokes since the start of 2018. The map is relative to the batsman who is positioned at the top of the rectangle in the center of the plot.*

Given how many contextual features are at play in cricket, it is paramount that a model has the capacity and ability to capture the "specificity" of the given situation (i.e. the identity of the batsman and bowler, the fielders and the game-state). Unfortunately, however, current methods do not capture these important contextual features and as such, the best analytics currently generated in cricket rely on either broad averages that ignore context or ever-diminishing sample size. An example is shown in Figure 2, where we show Ben Stokes' hitting chart for England which is an aggregate of his run scoring areas since the start of 2018 in One Day Internationals, but these do not consider the ball trajectory, bowler or match situation.

## 3. Related Work

In terms of cricket, no previous work on personalizing predictions on shot locations has been done before. Previous analyses have concentrated on scorecard level data for performance analysis, such as the rating of batsman performance in Test match [3] and One Day [4] forms of the game. Others have looked to simulate match scores [5] or predict optimal run scoring strategies [6], none of which utilize the spatial or shot type data to aid in team strategies.

A comparison could be made with the "field shift" in baseball, when fielder locations change from the "traditional" starting points [7] to counter batter hitting trends. Shot locations from spray charts would be used to observe players who had a strong preference for certain areas of the field. For example, as of 2015, the Tampa Bay Rays used a regression model that predicted the likelihood of batters hitting the ball into various infield zones and would shift if any zone had over 40% ground balls [8]. This was improved upon by using mixed-effects models which could predict the likelihood of ground balls in 9 zones using league wide fixed effects and random effects to account for individual player tendencies [8]. In tennis, a K-SVD method of learning player style from trajectories to predict the outcome of a point has been proposed [9].

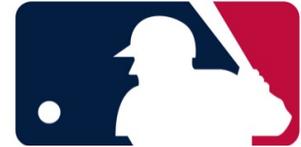

## 4. Methods

To reach our goal of revolutionising predictions of matchups between bowlers and batsmen in cricket, we first require a definition of shot type by a batsman. As of today, analysis of batsman shot type is limited and typically based on the end location of each shot. The drawback here is that the distance the ball travels is highly dependent on the location of the fielders who intercept the ball. Therefore, shot direction is the main information teams take from the data since ball by ball fielder location data is rarely collected. Hence, raw analysis of shot end locations obscures the intent of the batsman which leaves us to speculate whether the fielders should be positioned close or on the boundary. In shorter forms of cricket, there is a restriction on the number of fielders on the boundary too, so optimal placement of these players is a necessity. Consequently, to better understand batsman shot preference we propose analysis of shot intent and direction rather than end location alone.

We developed a personalized deep learning approach to predict the likelihood of specific shot types for any given delivery, trained using Opta ball-by-ball data from the past 8 years of international cricket (over 430,000 balls). Opta's highly detailed dataset labels each shot with one of 25 shot types, which for our purposes we assign a label of 0, 1 or 2 for increasing aggression. These aggression designations were based on exploratory analysis of every shot label and corresponding runs scored. We combine our aggression labels with the shot angle to create bespoke target variables based on splitting the field into 16 zones that follow standard cricketing nomenclature. We also include a defensive zone for when the ball is not hit with any aggression (aggression label 0), resulting in 17 target variables in total. The target variable representation is demonstrated in Figure 3 below. These zones effectively measure intent and shot angle rather than where the ball is fielded. This provides a clearer description on where the batsman is attempting to hit the ball and is therefore a superior guide for fielder placement and bowling tactics. One should note that the third man and fine leg zones near the top of the field combine both aggression labels. This is because the direction of the delivery by the bowler is already towards these zones, so shots played up the batsman usually aim to deflect the ball and take advantage of its natural velocity. As a result, such shots will usually reach the boundary in this direction if no fielders are positioned to intercept it.

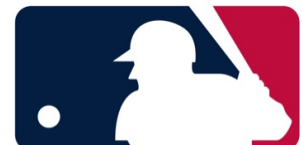

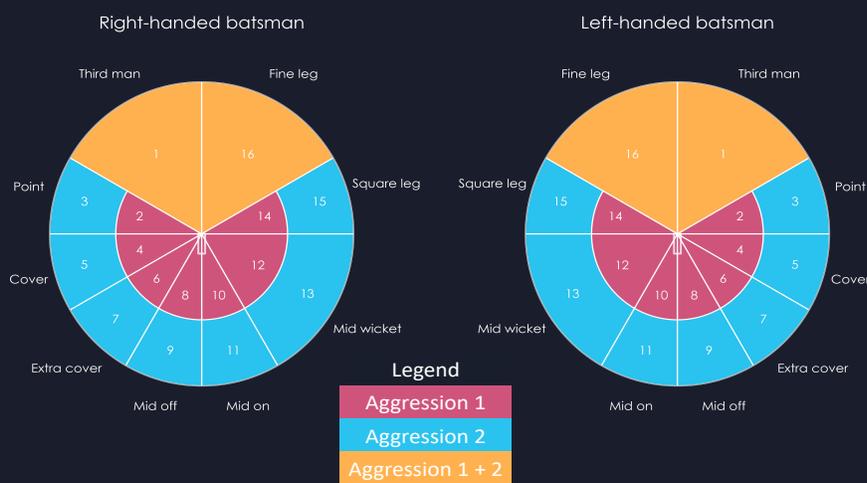

*Figure 3: Aggression labels for the corresponding raw shot label in the data set, with total number of deliveries falling into each category. Lower: Spatial map of the bespoke target variables.*

To predict shot type, we develop a deep learning model that consists of two main components to make use of the various feature types. We utilize a multi-layered long short-term memory (LSTM) recurrent neural network for our ball-by-ball data, which contains match context information in addition to delivery trajectories. We augment this LSTM with a multi-layered feed forward neural network containing personalized information for the batsman and bowler of each delivery, based on their historic data.

### 4.1. Shot Prediction Using Match Features

Cricket matches consist of a sequence of 1v1 events, for which we have an abundance of detailed information. Our data consists of ball-by-ball delivery information such as, line and length (where the ball lands on the pitch), movement of the ball both through the air and off the pitch (swing through the air or spin direction after bouncing), handedness of the bowler, style of bowler (spin vs speed), as well as the angle from which they deliver the ball relative to the wickets at the bowlers' end (referred to as "over the wicket" or "around the wicket" in cricketing vernacular). These bowling angles along with the line and length zones are defined in Figure 10, where the bowler lets go of the ball at the "popping crease" and the direction of travel would be towards the top of the plot.

By itself, this information can provide useful insight on the likely shot type; different line and length combinations increase the likelihood of success for certain types and reduce the likelihood of others. However, the probability of shot success is only a single factor a batsman will take into account when making their split-second decision on shot type.

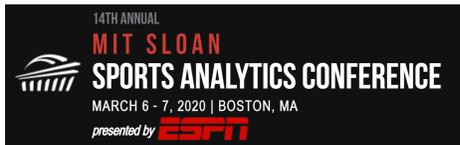 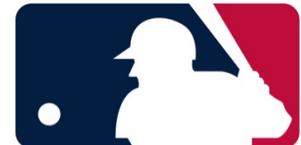

Another key factor is the current match situation. For example, if a batsman has faced few deliveries in the match, then safer shot types are often the preferred option until they get acquainted to the speed the ball bounces from the pitch and atmospheric conditions which can influence ball movement through the air. On the other end of the scale, after a batsman has established themselves by facing many deliveries, then aggressive shot intent is more likely. All of these decisions may also depend on the period of the team's innings, the number of wickets the team has left and current field placement restrictions.

To this end, we supplement the ball-by-ball delivery information with ball-by-ball match context features, including team information such as stage of the innings, wickets taken by the bowling team and runs scored by the batting team. We also add information specific to the batsman, such as their current runs scored, and deliveries faced. These match features add context to the delivery trajectory information to provide a more detailed description of factors that will influence the batsman's choice of shot type.

Our initial analyses made use of memoryless models using delivery information and match context features to predict the shot type across the 17 zones. This totalled 39 features which we identified as having the most influence on shot decisions from the huge quantity of potential contextual variables. First, we used a random forest approach which easily surpassed the naïve model which predicts a constant shot likelihood based on their global proportions. Accuracy almost doubled from 10.9% using this naïve approach, to 19.9% for the random forest on our holdout test set of 86,941 deliveries. Meanwhile, the log loss was improved from 2.83 in the naïve model to 2.51 using the random forest. We saw further gains when developing a feed forward neural network model which again improved our predictive capability to 20.3% accuracy and a log loss of 2.48. These results are reported in Table 1.

Although these numbers showed significant performance gains, the ball-by-ball nature of cricket lends itself to time series analysis. To make use of the time series information, we modelled both the match context and delivery information using a multi-layered LSTM with a lookback window of 6 (including the current delivery); this matches the number of deliveries per over within a cricket match. Although this makes sense within the format of the sport, this lookback value was chosen to provide the best model performance. Our LSTM network provided further gains over the memoryless alternatives, increasing the accuracy to 21.0% and the log loss to 2.45 (Table 1).

### 4.2. Adding Personalized Features
The match and delivery information features provide context for the batsman to make their shot decision. However, clearly the final shot type depends on the batsman themselves - their personal preference and ability. Shot type can be broken down into multiple levels. Some players will prefer to work the ball around the field to steadily accumulate runs (1, 2 run shots) throughout their innings, whilst others will look for big shots (4, 6 run shots) to score more quickly. In addition to this, different batsmen prefer to target certain areas of the field; some are stronger hitting straight for example, whilst others prefer hitting at 90-degree angles.

For this reason, we include features utilising personalized information on both the batsman and bowler in a similar fashion to previous studies [10, 11]. Our personalized batman features include measures of ability and aggression for various delivery trajectories, whilst also providing general information on the batsman's favored hitting directions. We also include bowler information such as the average number of runs scored, proportion of dot balls (0 runs) and boundaries (4s or 6s) for different delivery trajectories.

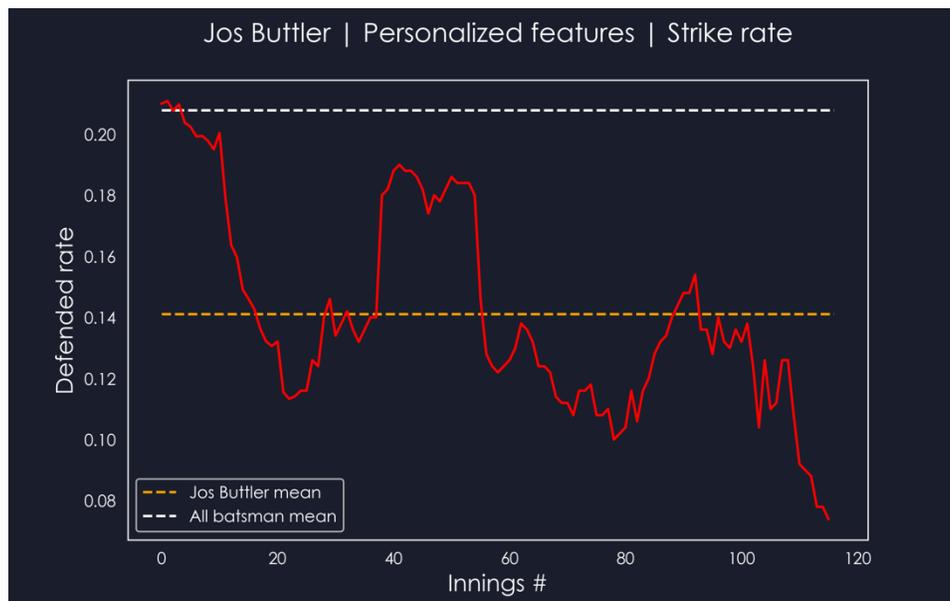

*Figure 4: Example of an updating personalized embedding for proportion of defended shots by England's Jos Buttler.*

To ensure that our personalized features are dynamic and account for changes in player ability and preferences over time, we use data from the previous 500 deliveries that each player has faced. This also allows predictions to be based on the most relevant and up-to-date information possible. For players who have faced less than 500 deliveries (approximately 10% of deliveries in our data set have at least one of the batsman or bowler with less than 500 deliveries), we use a linearly weighted average between the player's value and the global average value for that feature, based on how many deliveries the player has participated in. For example, a player with only 100 deliveries before a match would see their personal historic data contribute 20% to their features, with the global average contributing 80%. We demonstrate this with an example in Figure 4. This figure shows how England's Jos Buttler defensive shot rate decreases away from the global mean initially as he accumulates more data, before settling on the moving average after his 22nd innings when he has faced 500 deliveries in total. This ensures a steady move away from the `typical` player feature set to a personalized feature set as more data, and hence knowledge, is accumulated. In the Appendix, we visualize the high-dimensional personalisation space using t-SNE for both bowlers and batsmen in Figure 11 and Figure 12.

In total we have 46 personalized features, which we add as auxiliary variables to the output of the multi-level LSTM trained on delivery and match context features. We then run this through a 2-level feed forward neural network before the categorical predictions to the 17 output groups (see Figure 5). This framework enables the model to combine personalized descriptive metrics with contextual time series information.

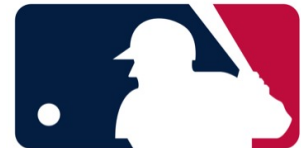

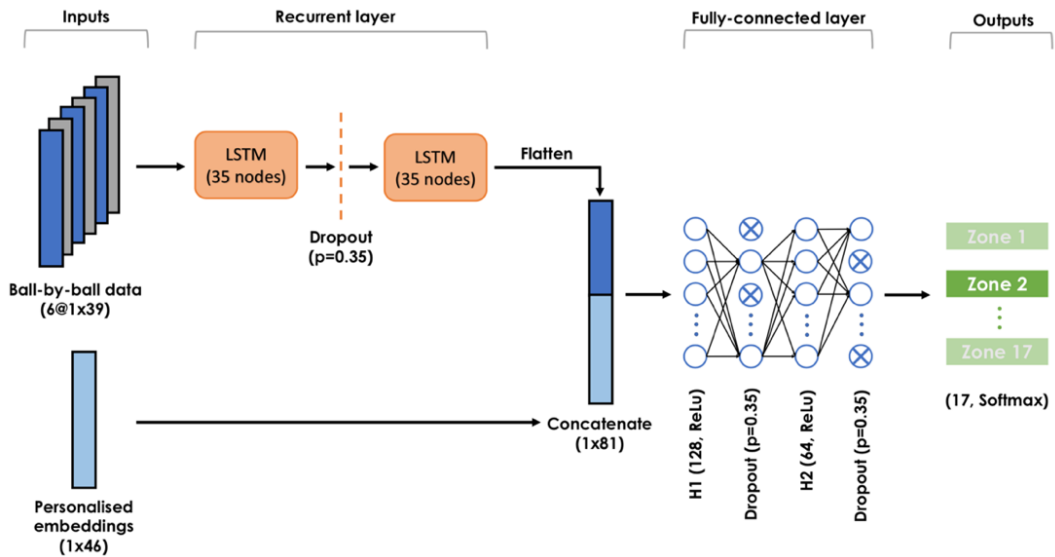

*Figure 5: Neural Network Architecture*

Performance was again improved over the previous model, with the accuracy increased from 21.0% in the non-personalized LSTM, to 22.1% in the personalized model. The loss was also improved from 2.45 to 2.42 (Table 1). Considering the number of zones and difficulty of exact shot prediction over such a wide area, we believe that these results show considerable improvement and value over our less detailed models.

*Table 1 Prediction accuracy and log loss over 17 shot zones on the hold out test set of 86,941 deliveries.*

| Model | Accuracy (17 Zones) | Log Loss |
|---|---|---|
| Naïve | 10.9% | 2.83 |
| Random Forest | 19.9% | 2.51 |
| Feed Forward Neural Network | 20.3% | 2.48 |
| Long Short-Term Memory Network | 21.0% | 2.45 |
| Long Short-Term Memory Network - Personalized | 22.1% | 2.42 |

# 5. Personalized Prediction Examples

To demonstrate the outputs of our model, we refer back to the introduction surrounding the final of the 2019 World Cup in England. We split this across three major elements:

1. Simulating personalized batsman predictions
2. Pre-match tactical planning to optimize batsman-bowler matchups and field placements
3. In-game tactics tailored to the on-going match context

## 5.1. Personalized Batsman Predictions

To illustrate the value and impact of including batsman-specific personalized features, we simulate the same set of deliveries from a given bowler during the final for a number of different batsmen. As such, the contextual, sequential and personalized-bowler features are fixed, which means any differences in the predictions are purely due to the different batsman used.

We use deliveries from the left-arm pace bowler, Trent Boult of New Zealand, who is one of the world's leading bowlers, versus the top 6 batsmen on the England cricket team. These batsmen make up one of the best batting units in ODI history and are made up of a mix of four right-handed batsman (Jason Roy, Jonny Bairstow, Joe Root, Jos Buttler) and two left-handed ones (Ben Stokes, Eoin Morgan).

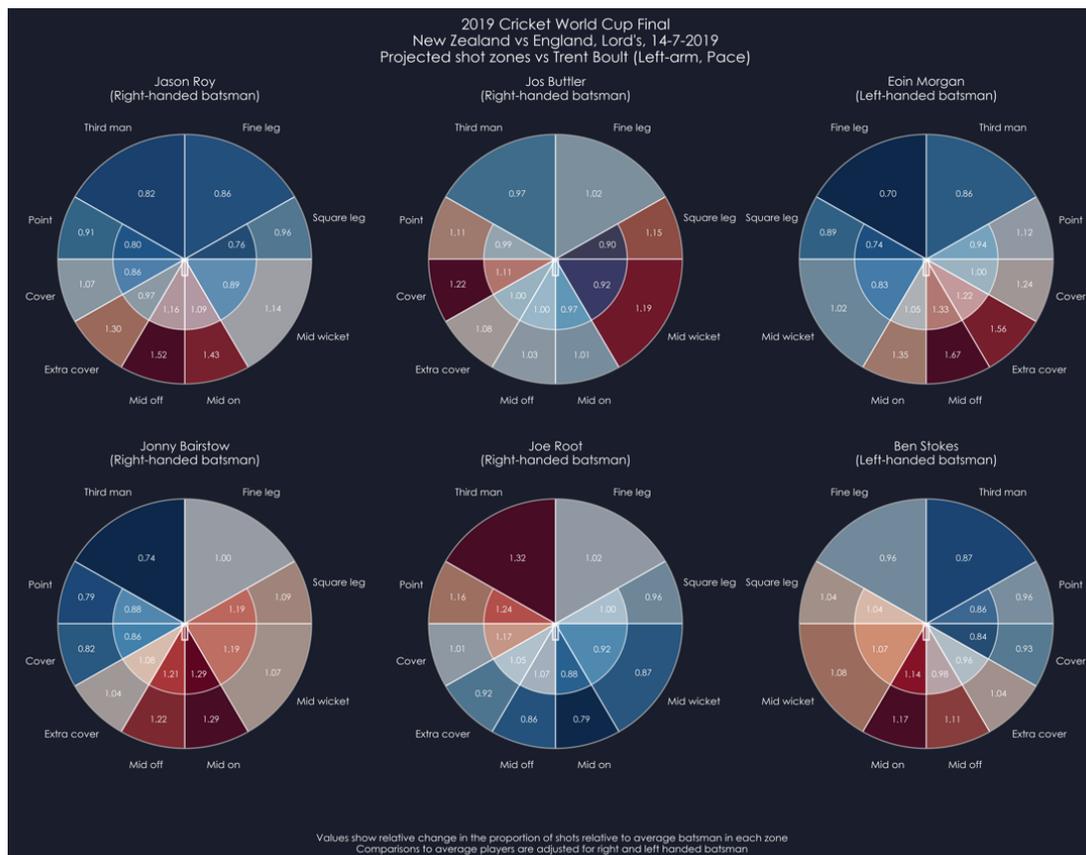

*Figure 6: Projected shot zones for England's top-six batsman when facing deliveries from the New Zealand bowler Trent Boult in the 2019 Cricket World Cup final.*

Taking the first column of Figure 6 above, the differences are relatively subtle but do point towards the batsman's own preferences in terms of both angle and aggression. The proportion of aggressive shots by each batsman is larger in the case of Roy relative to Bairstow (50% vs 43%), particularly through the mid off region. Both batsmen strongly favor aggressive straight shots compared to the average batsman, while being less likely to aim their shots through the third man-point-cover arc. Bairstow's less aggressive strokes also favor the straighter zones (mid off & mid on) in addition to the square leg and mid wicket regions.

On the second column of Figure 6, two more right-handed players are shown; the pair are arguably England's two premier batsman – albeit possessing very different traits. In terms of aggression, Root is the least forceful of England's top 6 batsman with 40% of his shots classified as aggressive compared to 47% for Buttler. A striking difference between Root relative to other right-handed batsmen and his England teammates shown here is the proportion of his shots directed through the third man zone. The natural angle of a left-handed bowler to a right-handed batsman from over the wicket (which is broadly the case here) is for the ball to travel "across" the batsman, opening up the potential to deftly deflect the ball into a zone that typically has only one fielder patrolling it. Buttler shows a tendency to strike the ball aggressively towards either side of the field compared to a typical batsman, while still being close to the average shot proportion through the straighter zones. This contrasts with his colleagues who clearly diverge from the average batsman in certain areas, supporting his reputation as a "360 degree" player.

On the third column of Figure 6 two left-handed batsman are shown (note that the field location labels are flipped, as explained in Figure 10). Firstly, Stokes is a less aggressive player than Morgan, with 42% of his shots being classed as aggressive vs 48%. In terms of shot direction, the natural angle of a left-handed bowler to a left-handed batsman would generally dictate the batsman playing more shots through the arc from fine leg to mid wicket. Stokes follows this general expectation as he attempts a slightly above average amount of his shots through these zones, although with a clear preference for shots straight down the ground. Morgan however is far less likely to target the most-favored zones by left-handed to these deliveries, instead heavily favoring the mid off and extra cover regions to the extent that he is more predictable in his shot types compared to his colleagues.

What the above exploration and examples demonstrate is the differing make-up of England's top batsman compared to each other and the typical batsman. There is a balance between more and less aggressive players (Roy/Morgan/Buttler and Bairstow/Root/Stokes respectively), which aligns with their reputations. Aside from the opening pair of Roy and Bairstow, England operate a "fluid" batting order which can change depending on match context. In a less favorable situation, Root or Stokes will be deployed to consolidate the innings and mitigate the risk of a low score; when aggressive stroke-play and accelerated scoring-rate is required, Morgan or Buttler will be sent to the crease earlier than their nominal batting positions of fourth and sixth. Our predicted shot zones also illustrate clear differences between the batsmen that is a consequence of not only their handedness, but their favored hitting zones; again, England are known to consider such factors when choosing their batting order if the opposition bowlers being played or dimensions of the venue favor one over the other.

### 5.2. Pre-match tactical planning

With the advent of video and data analysis in cricket, the level of available pre-game information and potential scope to devise tactical strategies for each opponent has sky-rocketed. Our personalized predictions are ideal for such a task as we can isolate both individual batsman and the stage of their innings and then simulate a suite of parameters to study their performance.

For our example, we will focus on New Zealand's potential planning against the England batsman Ben Stokes. We do this by drawing from Stokes' innings in the previous year leading up to the final in order to find the typical match context when he is batting (i.e. the match score and stage of the innings). We then use this match context information and explore a large parameter space of four different bowling lines, four different bowling lengths and five different New Zealand bowlers who were selected for the final. For the four right-handed bowlers, we also vary the side of the stumps from which they deliver the ball; a left-handed pace bowler facing a left-handed batsman will practically always bowl from the same side.

In total, this parameter space spans 144 different combinations, which would be impossible to explore through standard data analysis due to sample size restrictions.

In Figure 7 we illustrate this by exploring the shot predictions of Ben Stokes at the start of his innings (0-9 balls) when facing different delivery lengths (see Figure 10-B), with all other parameters fixed. We aggregate his shot types into aggression levels and direction in order to summarize the outputs, although we could explore the full detail of the model predictions.

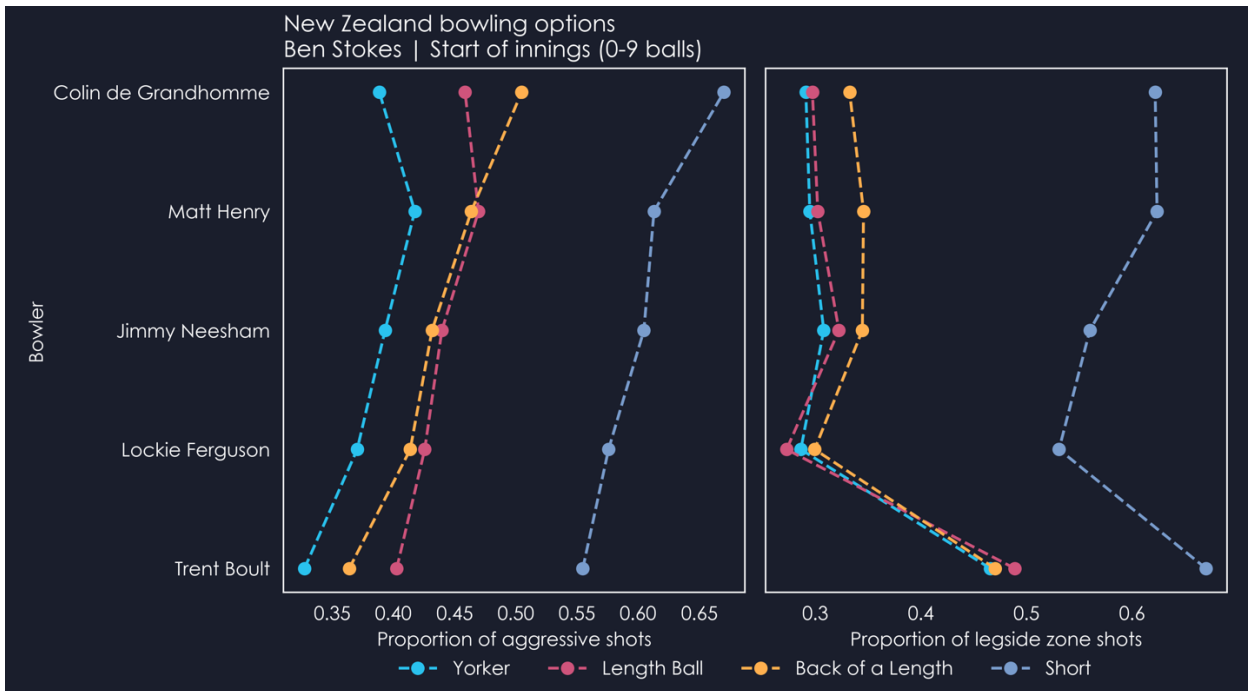

*Figure 7: Simulated shot types by Ben Stokes when facing different New Zealand pace bowlers at the start of his innings (0-9 balls faced) against varying delivery lengths. Note that Trent Boult bowls left-handed, while the other bowlers are right-handed. Bowlers are ordered by aggressive shot proportion.*

Firstly, we can observe that Stokes will generally play more aggressive shots when facing Colin de Grandhomme and Matt Henry – two of the slower-paced New Zealand bowlers. He is predicted to play the least aggressive shots when facing the New Zealand left-arm fast bowler, Trent Boult. In terms of the direction of his shots, Stokes generally favors the off side of the pitch (i.e. shots towards his left-hand side), except for against Boult when the balance is more even between that and the leg side.

In terms of length of delivery and attacking intent and direction, the predictions clearly demonstrate that Stokes will play far more aggressive shots against the short deliveries and that these will be more biased towards the leg side. His attacking intent against short and "back of a length" balls is particularly enhanced against de Grandhomme compared to the other bowlers – de Grandhomme bowls at a relatively low speed for international cricket (<130 km/hr) and balls bowled short at such speeds are much easier to hit.

In cricket, batsmen generally go through a period of "playing themselves in" as they adjust to the bowlers and to the pace of the pitch, and at these times are more vulnerable to being dismissed and typically score at a slower pace. To illustrate this, we explore the same parameter space as above but now we also predict against cases where Stokes has faced more than 60 balls. In Figure 8 we compare our predictions at different stages of Stokes' innings to observe changes in his predicted behavior, using the same aggregated metrics as in Figure 7.

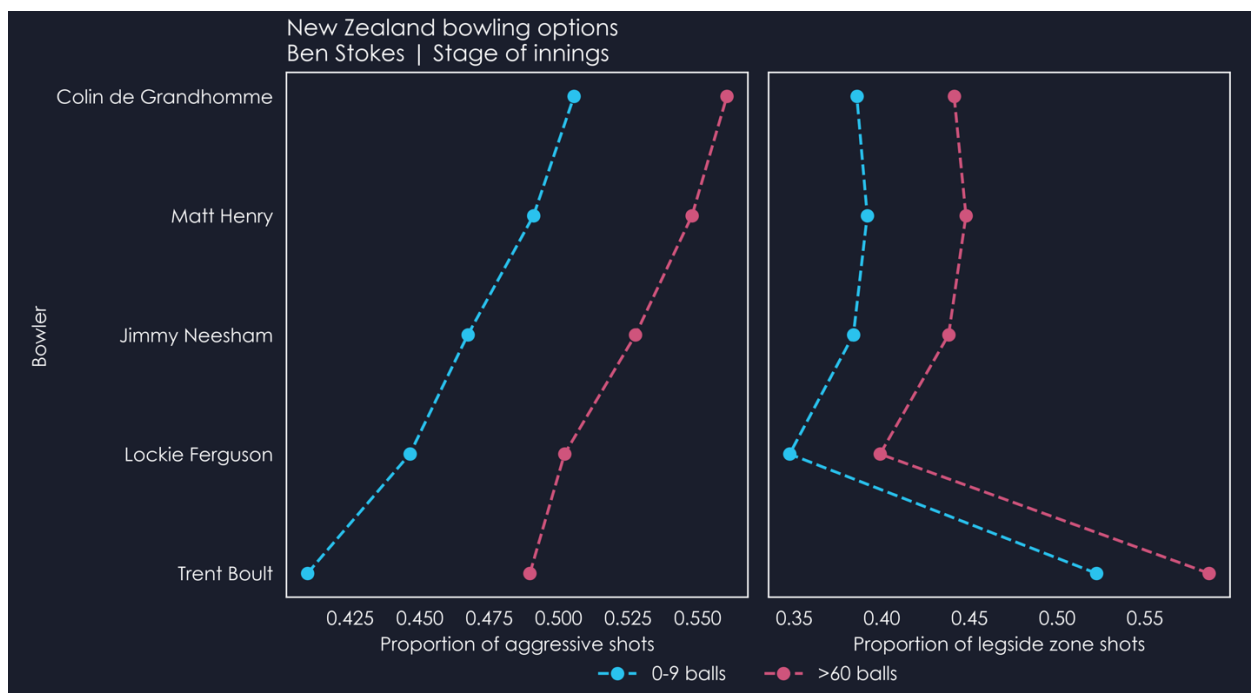

*Figure 8: Simulated shot types by Ben Stokes when facing different New Zealand bowlers at the start of his innings (0-9 balls faced) and later in an innings (>60 balls faced). Note that Trent Boult bowls left-handed, while the other bowlers are right-handed. Bowlers are ordered by aggressive shot proportion.*

When comparing the predictions in this manner, we observe clear differences in both Stokes' attacking intent and the direction in which he plays his shots. This is the case against all of the bowlers, although the change is not completely linear; his aggressive shot proportion increases by 5-6% against the right-handed bowlers, while it increases by 8% against Boult as he appears to be more cautious against him early on.

What the above results illustrate is that we can derive clear signal from our predictions to inform pre-match planning that can be delivered across a batsman's innings. Such insights could then be utilized in-game for both bowler selection and field placements.

### 5.3. In-game tactics
To illustrate the power of our predictive model for in-game tactics, we return to our motivating example from the introduction, with England requiring 9 runs to win from the final 3 deliveries of the 2019 World Cup final.

As already demonstrated, the length of the delivery strongly alters the predicted shot type and zone targeted by England's batsman Ben Stokes in this critical moment.

We can also illustrate how valuable our personalized bowler features can be for shot prediction by swapping out the left-arm pace bowler of Trent Boult, for New Zealand's right-arm pace bowler Lockie Ferguson. In the actual match, Ferguson had bowled his maximum allotted deliveries by the time this vital delivery was bowled, but we can visualize how different Stokes' predicted shot map would have been for the same line and length delivery if Ferguson had the opportunity to bowl it.

The top panels in Figure 9 show the shot map for Ben Stokes vs Lockie Ferguson for the yorker length and off stump line, the same line and length which was attempted by Trent Boult in the actual match. We include 2 plots for this line and length combination, one shot prediction for a delivery bowled by Ferguson over the wicket and one from around the wicket. The right-handed Ferguson bowling from around the wicket bowls from a similar angle to the left-arm Boult, whilst over the wicket would be from a significantly different angle (refer to Figure 10-B). Therefore, the closest comparison to Boult's attempted delivery (yorker, off stump; middle plot of Figure 1) by Ferguson is the top right plot of Figure 9. We can clearly see the difference in Stokes' likely shot direction due to the personalized bowler information used in our model.

Further to this, comparison of the shot predictions for over/around the wicket in Figure 9 show more details that the model picks out which can provide valuable predictive information for coaches and captains. We can observe how the difference between Stokes' shots vs Ferguson for over the wicket and around the wicket deliveries (left and right column respectively) vary depending on the length. For very full-length deliveries (which bounce near the batsman), the model shows that the angle makes little difference. However, once the length of the delivery is moved shorter to "length-ball" we see a huge change in shot probabilities, with "over the wicket" resulting in a far less concentrated shot area.

These examples show how our deep learning model captures the dynamic relationship between both the batsman and bowler abilities, whilst also accounting for how these interact with the trajectory information of the delivery. The interactions can be used to help teams develop bowling tactics, both in terms of delivery types to bowl and the bowler to bowl it, to suit the match situation and batsman they are bowling to.

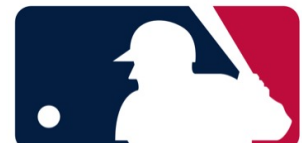
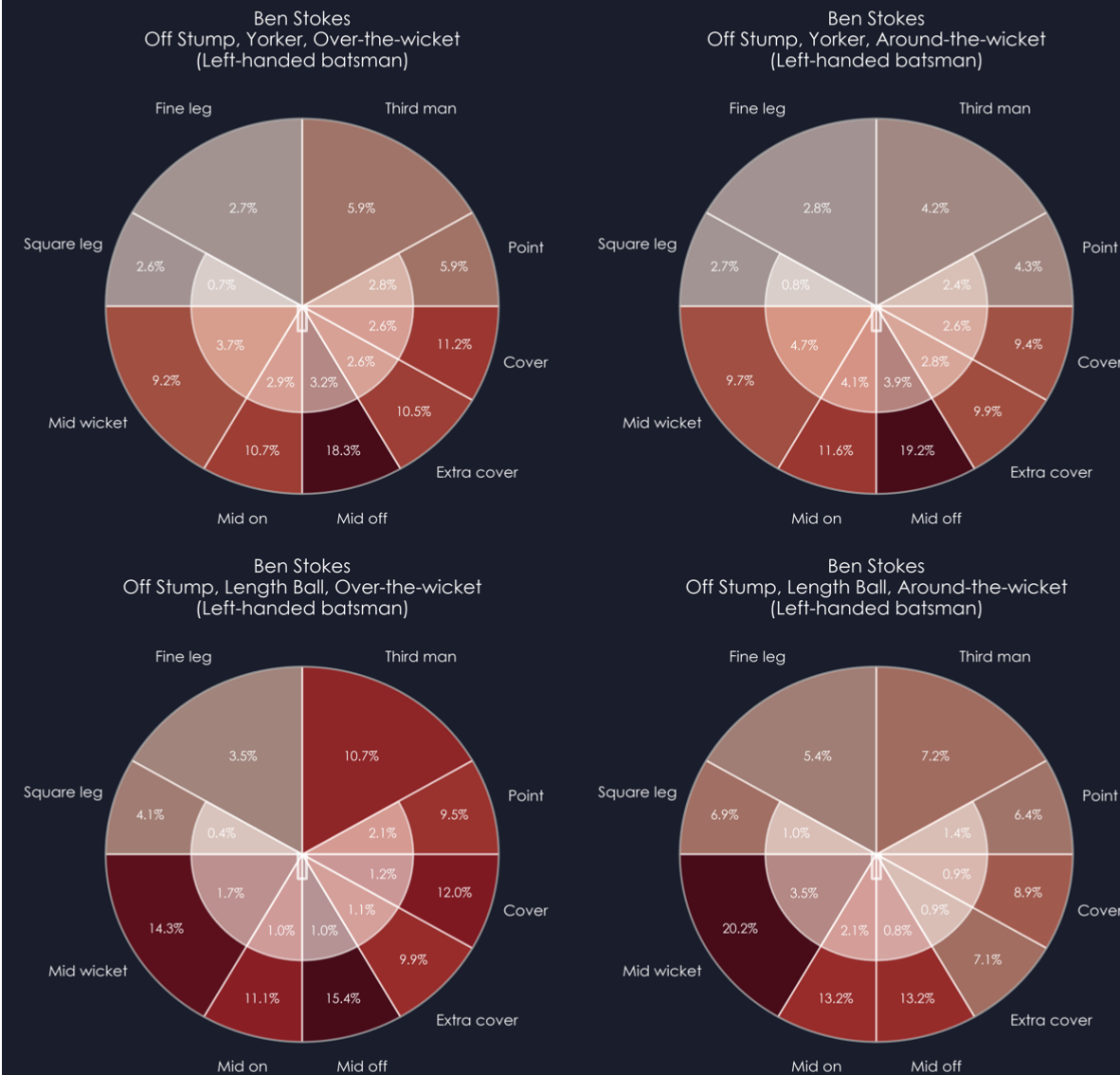

Figure 9: Projected shot zones for an alternative bowling selection for New Zealand (Lockie Ferguson) and a range of simulated trajectories for the 4th ball of the 49th over of the 2nd innings of the 2019 Cricket World Cup final. Ben Stokes (England) was the bat

# 6. Discussion

In this paper we have proposed a unique personalized deep neural network model, which utilizes our bespoke shot type definitions to predict likely batsman shot decisions. The power of our personalized predictions is that rather than relying on either broad averages that ignore context or ever-diminishing sample sizes, we can isolate and account for match context, the sequential nature of cricket and the individual tendencies of batsmen and bowlers. As a result, our predictions can add value to both pre-match line-up decisions that are most suitable to the opposition and venue, as well as in-game tactics.

Concerning our example of Trent Boult's bowling tactics during the 2019 World Cup final, it is interesting to postulate that his strategy became somewhat predictable. Having landed two yorker deliveries outside off stump that Stokes was unable to hit with any force – the third delivery seemingly aimed to repeat the trick but failed for two reasons as Boult bowled the ball fractionally too short and Stokes appeared to premeditate the delivery and position himself to play an extraordinary "slog-sweep" shot through mid wicket that landed halfway up the grounds' seating stand. A well-worn cricketing maxim was that landing six yorkers in the closing over was the optimal strategy for a bowler due to their difficulty to be struck cleanly by the batsman; such received wisdom ignores two important factors:

  i. The margin for error is incredibly small as a ball either slightly too full or too short can be easily dispatched to the boundary – even the best bowlers cannot always hit the same spot.
  ii. In an age of innovation and targeted practice, many batsmen are now capable of adjusting their approach to strike such deliveries cleanly, especially when the bowler's plan is fairly predictable

A more varied strategy from Boult may have improved New Zealand's chances, something which the tailored predictions of our model could have aided. By assessing the likely shot type for various delivery trajectories dependent on the current match situation, decisions could be made on which attempted delivery and fielder location combinations offer the best opportunity to restrict the batsman and also which are the least dangerous combination in the event a delivery is executed badly.

Ultimately, cricket is a sport often decided by the barest of margins: just a single run saved by New Zealand across their 300 deliveries would have been enough to see them win the World Cup. Our personalized predictions can provide vital information to inform the decision-making of coaches and captains, both in terms of pre-match and in-game tactical choices.

# Appendix

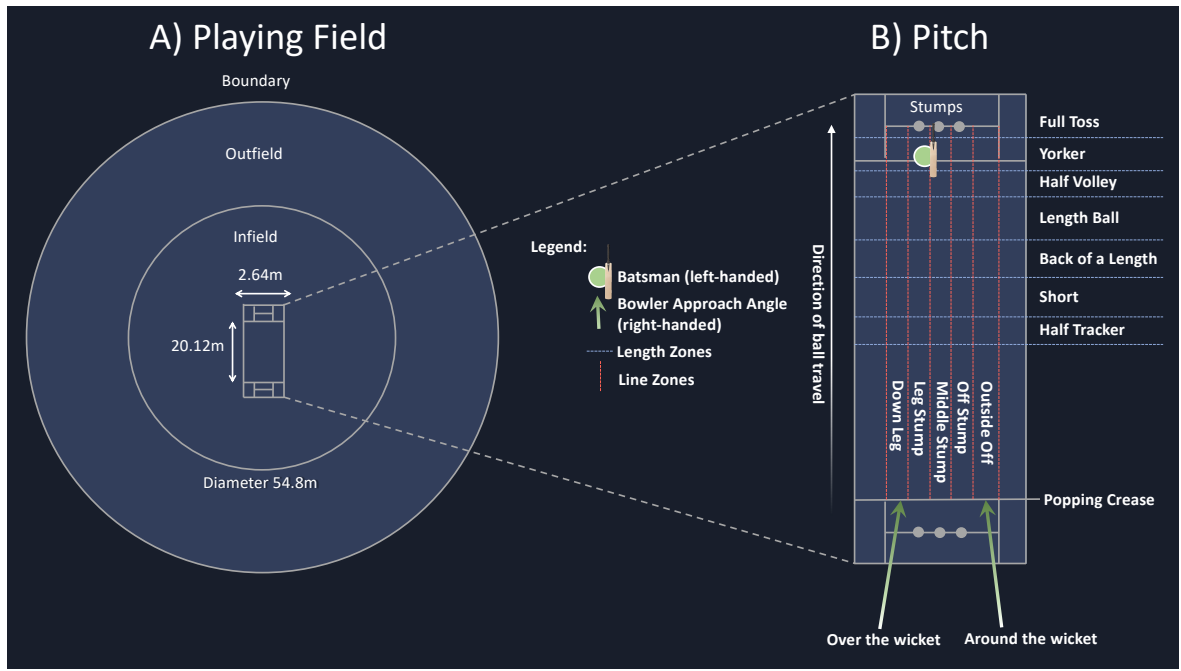

Figure 10: A) Map of playing area in cricket. The International Cricket Council stipulate that the boundaries of the field should be between 59.4-82.3 m from the centre of the pitch, equating to an area of 11,000-21,000 m$^2$ [12].
B) Pitch map for various line/length zones and bowling angles. Note that these line zones are for a left-handed batsman and would be mirrored on the middle stump for a right-handed batsman. The bowler approach angle names are for a right-handed bowler and would be swapped for a left-handed bowler.

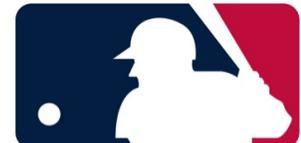
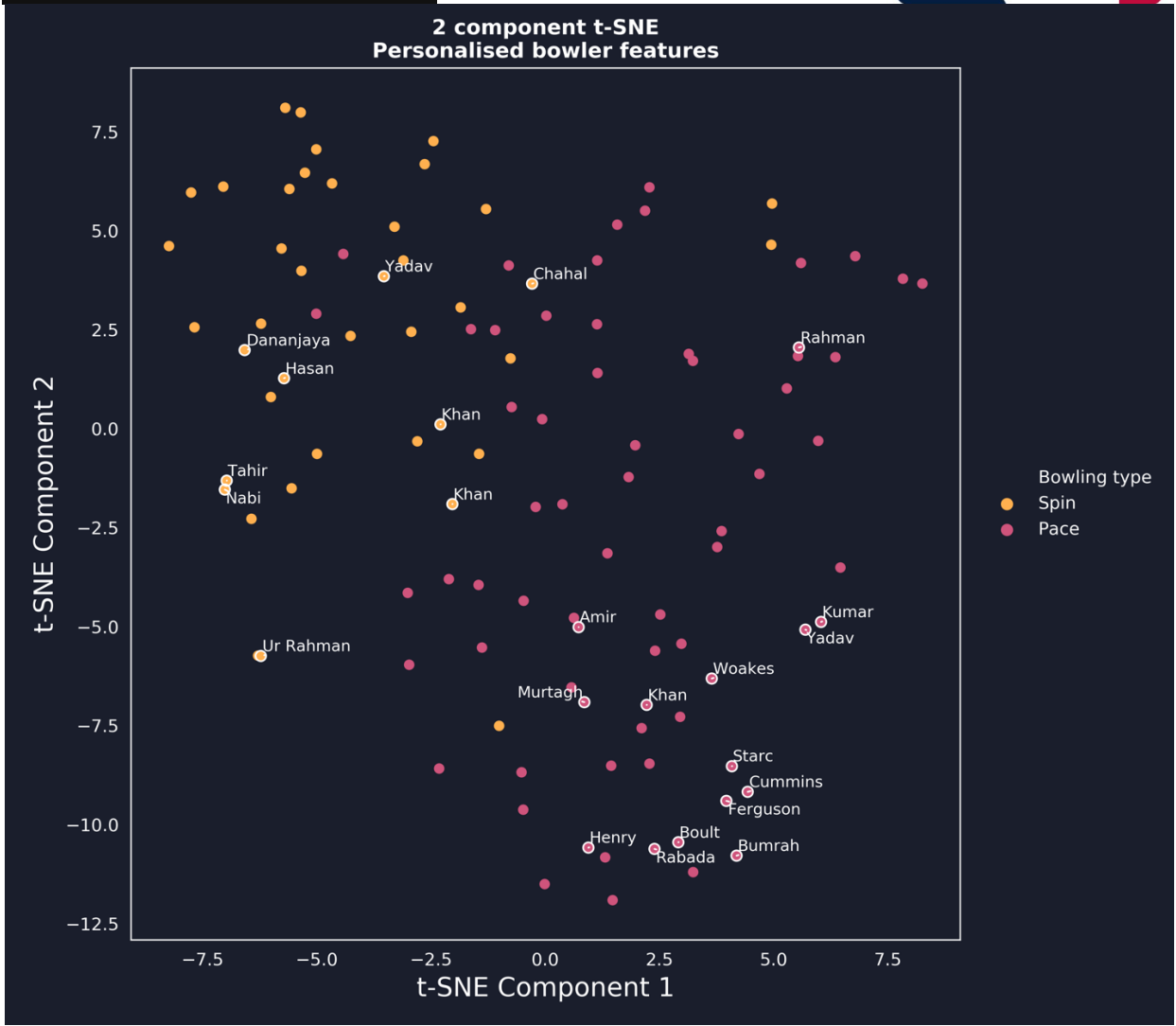

*Figure 11: A 2-dimensional t-SNE representation of the bowler feature space. Data points are each bowler's most recent set of features for players who've played more than 10 games in our dataset. Note: No distinctive spin or pace features were used in our feature space.*

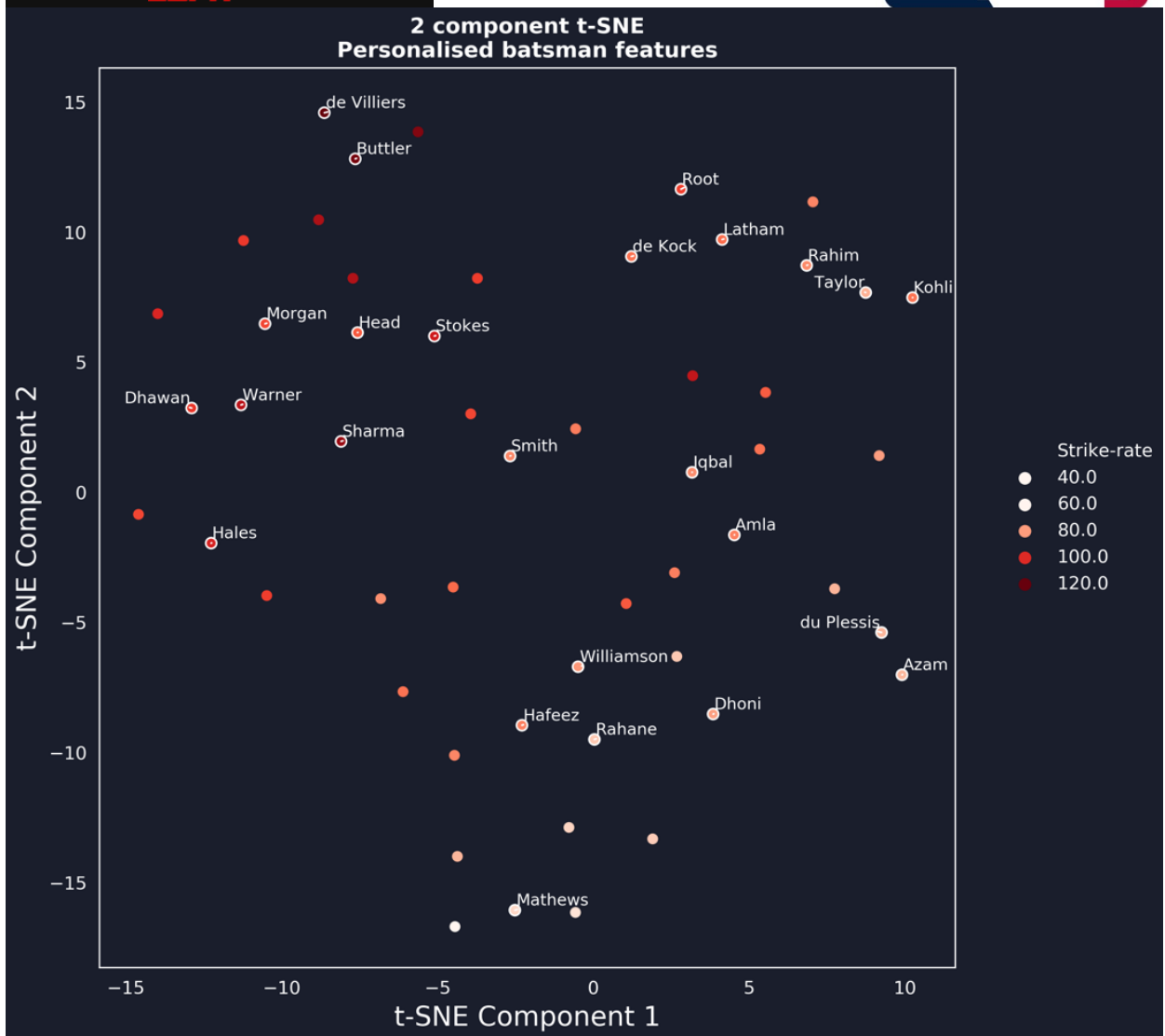

*Figure 12: A 2-dimensional t-SNE representation of the batsman feature space colored by strike-rate (average runs per 100 deliveries). Similar players are grouped together such as big 360-degree hitters de Villiers and Buttler.*

20